\pdfoutput=1
\documentclass[showpacs,superscriptaddress,aps,pra, twocolumn]{revtex4-1}
\usepackage{caption}
\usepackage{subcaption}
\usepackage{cancel}
\usepackage{amsfonts}
\usepackage{amsmath,amsthm}
\usepackage{bm}
\usepackage{amssymb}
\usepackage{mathrsfs}
\usepackage{amsmath, amsthm, amssymb,amsfonts,mathbbol,amstext}
\usepackage{mathtools}
\usepackage[colorlinks=true,citecolor=blue,urlcolor=blue]{hyperref}
\usepackage[normalem]{ulem}
\usepackage{mathrsfs}
\usepackage{dsfont}
\usepackage{graphicx}

\newtheorem{thm}{Theorem}
\newtheorem{cor}[thm]{Corollary}

\def\be{\begin{equation}}
\def\ee{\end{equation}}
\newcommand{\mc}[1]{\mathcal{#1}}
\usepackage{color}
\definecolor{violeta}{cmyk}{0.07,0.90,0,0.34}

\usepackage{cancel}
\usepackage{soul}
\definecolor{cgreen}{RGB}{26, 199, 76}

\begin{document}


\title{Graph Approach to Extended Contextuality}

\author{Barbara Amaral}
\affiliation{Departamento de F\'isica e Matem\'atica, CAP - Universidade Federal de S\~ao Jo\~ao del-Rei, 36.420-000, Ouro Branco, MG, Brazil} 
\affiliation{International Institute of Physics, Federal University of Rio Grande do Norte, 59078-970, P. O. Box 1613, Natal, Brazil}

\author{Cristhiano Duarte} 
\affiliation{International Institute of Physics, Federal University of Rio 
Grande do Norte, 59078-970, P. O. Box 1613, Natal, Brazil}
\affiliation{Schmid College of Science and Technology, Chapman University, One 
University Drive, Orange, CA, 92866, USA}


\begin{abstract}
Exploring the graph 
approach, we restate the extended definition of 
noncontextuality provided by the contextuality-by-default framework.  This 
extended definition avoids the assumption of 
\emph{nondisturbance}, which states that whenever two contexts 
overlap,  the marginal distribution obtained for the intersection must 
be the same.  
We show how standard tools for characterizing contextuality can also be used 
in this extended framework for any set of measurements and, in addition, we also provide several conditions that 
can be tested directly in any 
 contextuality experiment. Our conditions reduce to traditional 
ones for noncontextuality if the nondisturbance assumption is satisfied. 
\end{abstract}

\maketitle

\section{Introduction}

Quantum theory provides a set of rules to predict probabilities of different outcomes in
different experimental settings. Quantum predictions match with extreme
accuracy the data from actually performed 
experiments~\cite{Delft-free-15,Viena-free-15,Delft-free-16,Tonhao12,BBdMHP,STSM02,LLLZ15}, nonetheless they exhibit 
some peculiar properties
 deviating from the usual probabilistic description of classical systems~\cite{Hardy93,Bell66,Bell64}. One of these 
``strange'' characteristics
is the phenomenon of {\em contextuality}, which says that there may be no 
global probability distribution over a set of measurements consistent with the quantum theory recovering the right prediction for each \textit{context}, i.e. subsets
of compatible measurements. \cite{Specker60,Bell66,KS67,Fine82,AB11}.

A fundamental consequence of contextuality is that the statistical predictions of quantum theory cannot be obtained from 
models where measurement outcomes reveal pre-existent properties that are 
independent on  other compatible measurements that are 
jointly performed~\cite{KS67,Bell66}. This  limitation is related to the existence 
of incompatible  measurements in quantum systems and  thus represents an  
intrinsically non-classical phenomenon.
Besides its importance for a  more fundamental understanding of many aspects of quantum theory \cite{NBAASBC13,Cabello13,Cabello13c,CSW14,ATC14,Amaral14}, contextuality has also been 
recognized as  a potential 
\emph{resource} for quantum computing, \cite{Raussendorf13, HWVE14,DGBR14}, random number certification \cite{UZZWYDDK13}, and several other 
information processing tasks in the specific case of space-like separated 
systems \cite{BCPSW13}.

As a consequence, experimental verifications of contextuality have  
attracted much atten\-tion \cite{HLBBR06, KZGKGCBR09, ARBC09, LLSLRWZ11, BCAFACTP13} over the past years. It is thus of utmost
importance to develop a robust theoretical framework for contextuality that 
can be efficiently applied to real experiments. In 
particular, it is important to design a sound theoretical machinery well-adapted to
cases in which the sets of measurements do not satisfy the assumption of 
\emph{nondisturbance}~\cite{NBAASBC13}. This strong assumption 
demands that whenever the intersection of two contexts is non-empty, then the 
marginal probability distributions for  the  intersection  must 
be the same, a restriction that will hardly be perfectly satisfied in real 
experiments.

  In Refs.~\cite{KDL15,DK16,DKC16,DK17,DCK17}, the authors propose an alternative definition of 
noncontextuality, called \emph{contextuality-by-default} or \emph{extended contextuality},  that can be applied to any set of measurements. In this alternative definition,  a set of measurements  is said to be noncontextual 
(in the extended sense) if there is 
an extended joint probability distribution  which is consistent with the joint 
distribution for each context and, in addition,  maximizes the probability of  two 
realizations of the same  measurement  in 
different contexts being equal. Such a treatment reduces to the 
traditional definition of noncontextuality if the nondisturbance property 
is satisfied and, in addition, it can be verified directly from  
experimental data. Finally, in Refs.~\cite{KDL15,KD16} the authors provide necessary and 
sufficient conditions for extended contextuality in a broad class of scenarios, 
namely the $n$-cycle scenarios.

Focusing on developing an experimentally-friendly and 
robust theoretical framework for addressing noncontextuality, this contribution
explores the graph approach to contextuality,
developed in Refs. \cite{CSW10,CSW14, AII06} and further  explored in Refs.
\cite{RDLTC14, AFLS15, AT17}, to rewrite the definition of extended contextuality in graph-theoretical terms.
To this end, from the compatibility graph $\mathrm{G}$ of 
a scenario $\Gamma$, we define another graph $\mathscr{G}$ 
--the \emph{extended compatibility graph} of the scenario-- and 
show that extended noncontextuality  is equivalent to 
noncontextuality in the traditional sense with respect to the extended 
graph $\mathscr{G}$.
Within this graph-theoretical perspective, the problem of characterizing 
extended noncontextuality reduces to  characterizing 
traditional noncontextuality for the scenario defined by $\mathscr{G}$, a 
difficult problem  for general graphs \cite{Pitowsky91, DL97, AII06, 
AII08}. Nevertheless, the graph-theoretic approach we employ here allows one to use several tools already
developed for the study of contextuality also in the the characterization of extended contextuality.

 We have organized the paper as follows: in Sec. \ref{sec:comp} we  review the definition of 
a compatibility scenario and of noncontextuality in the traditional sense. 
In Sec. \ref{sec:ext}, we review the definition of extended 
noncontextuality as given in Refs.~\cite{KDL15,DK16,DKC16,DK17,DCK17,ADO18}, stating it in graph-theoretical 
terms. We define the corresponding extended scenario and show that the notion of contextuality is equivalent to traditional contextuality
with respect to the extended scenario. In Sec. \ref{sec:tools} we review several tools for the 
characterization of contextuality that can also be applied to the extended framework. In Sec. \ref{sec:couplings} we discuss different approaches to extended contextuality and argue why we
believe the one proposed in Refs. \cite{KD16, KDL15} is more suitable. We finish
 this work with a discussion in Sec.~\ref{sec:discussion}.


\section{Compatibility scenarios}
\label{sec:comp}

 A \emph{compatibility scenario} is defined by a triple \begin{equation} \Gamma 
:=\left(X,\mathcal{C}, O \right),\end{equation} where $O$ is a finite set, $X$ is a 
finite set representing 
measurements in a physical system whose possible outcomes lie in $O$, 
and 
  $\mathcal{C}$ is a family of subsets  of $X$. 
The elements $C \in \mathcal{C}$ are called \emph{contexts} and 
encode the compatibility relations among the 
measurements in $X$, that is, each set $C \in \mathcal{C}$ consists 
of a maximal set of compatible, jointly measurable measurements
\cite{AB11, AT17book}.

Equivalently, the compatibility relations among 
the elements of $X$ can be represented by a hypergraph. 
The \emph{compatibility hypergraph} associated with a scenario $\left(X, \mathcal{C}, 
O \right)$ is  a hypergraph \begin{equation}\mathrm{H} := \left(X, \mathcal{C}\right) 
\end{equation}
whose vertices are 
the measurements  in $X$ and whose hyperedges are the contexts $C \in 
\mathcal{C}$.

When a particular set of compatible measurements in a context $C=\{x_1,x_2,...,x_{\vert 
C \vert } \} 
\in \mc{C}$ is performed jointly, a list $a=(a_1,a_2,...,a_{\vert C 
\vert })$ of outcomes in $
O^{C}:=O \times O \times ... \times O$
must be observed.
The probability of this list of outcomes, with respect to this specific context, is 
denoted by
\be p\left(a \vert C\right) \coloneqq p\left(a_1, \ldots , a_{|C|}\left|x_1, \ldots , x_{|C|}\right.\right).\ee
The collection of all these
 joint probability distributions 
is usually called~\cite{NPA08,JNPPW112011} a \emph{behavior} $\mathrm{B}$ for the scenario $\left(X, \mathcal{C}, O \right)$.

In an ideal situation, it is generally assumed that behaviors must satisfy the  
\textit{nondisturbance condition}~\cite{Amaral14}. Such condition says that whenever two contexts $C$ and 
$C'$ overlap,
the marginals for $C \cap C'$, computed either  from  the distribution for $C$ or from  the distribution for $C'$, must coincide.
The set of nondisturbing behaviors will be denoted by  $\mathcal{ND}\left(\Gamma\right)$.
 
In the hypothetical situation where all measurements
in $\mathrm{X}$ are compatible, \emph{i.e.} in the extreme situation where there is 
a unique context, it would be possible to define a
\emph{global probability distribution} 
\be p(a_1 a_2 ... a_{\vert X 
\vert} \vert x_1 x_2 ... x_{\vert X \vert})\label{eqglobal}\ee dictating the probability of outcomes $a_1 a_2 ... a_{\vert X 
\vert}$ in a joint 
 measurement involving all measurements in $X$. 
 
 It is in a 
less extreme situation that the 
concept of (non)contextuality plays its role, though.
A behavior $\mathrm{B}$ is 
\emph{noncontextual} whenever the probability distributions assigned by $\mathrm{B}$ 
to each context can be recovered as marginals from a global probability distribution 
$p\left(a_1a_2 \ldots a_{|X|}\vert x_1 x_2 ... x_{\vert X \vert}\right)$ \cite{Fine82, 
AB11}. The set of noncontextual behaviors will be denoted by  
$\mathcal{NC}\left(\Gamma\right)$.
Notice that a noncontextual behavior is necessarily nondisturbing, what left us with the following inclusion
\be \mathcal{NC}\left(\Gamma\right) \subset \mathcal{ND}\left(\Gamma\right).\ee

\begin{figure}[h!]
\centering
\includegraphics[scale=1]{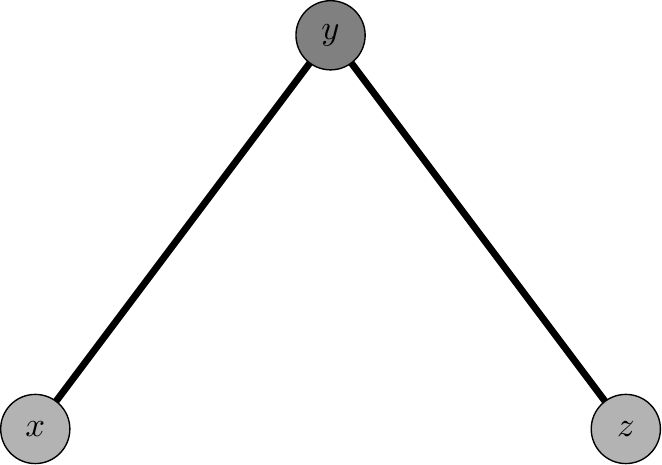}
 \caption{A compatibility scenario with three 
measurements $x,y,z$ 
and two contexts, $\left\{x,y\right\}$ and 
$\left\{y,z\right\}$.}
 \label{fig:c3}
\end{figure}

As an example, consider the scenario $\Gamma$ containing three dicotomic measurements $\{x,y,z\}$ where only one measurement, say $\{y\}$, is compatible with the two others. Mathematically, $\Gamma =\left(X, 
\mathcal{C}, O \right)$ with ${O}=\left\{-1,1\right\}$,   $\mathrm{X}=\left\{x, 
y, z\right\}$
and $\mathcal{C}=\left\{\{x,y\}, \{y,z\}\right\}$. The compatibility hypergraph of 
this scenario is a simple graph~\footnote{Mathematically a graph is said to be simple whenever there are no loops, double edges nor hyperedges}, as depicted in Fig. \ref{fig:c3}.

In this scenario, a behavior consists in specifying probability distributions
\begin{align}
 & p(ab|xy), \,\, a,b \in \{-1,1\} \\
 & p(bc|yz), \,\, b,c \in \{-1,1\}.
\end{align}
%
%
The nondisturbance condition demands that
\be p\left(b \vert y\right) \coloneqq \sum_a  p\left(ab\vert xy\right) = \sum_c 
p\left(bc \vert yz\right). \label{eq:ExampleNonDisturb}\ee
A behavior  is noncontextual if there is a global probability 
distribution $p\left(abc\vert xyz\right)$ such that 
\begin{align}
 p\left(ab\vert xy\right)&= \sum_c p\left(abc\vert xyz\right)\\
 p\left(bc \vert yz\right)&=\sum_a p\left(abc\vert xyz\right).
\end{align}


\section{Extended Contextuality}
\label{sec:ext}

To define noncontextuality in a scenario where the nondisturbance property 
 does not hold true, we shall first  consider \emph{extended global 
probability distributions} of the form 
\begin{widetext}
\be \label{eqextendedglobal}
p\left(\underbrace{a^1_{1} \ldots a^1_{\left|C_1\right|}}_{C_1} \underbrace{a^2_{1} \ldots a^2_{\left|C_2\right|}}_{C_2} \ldots
\underbrace{a^m_{1} \ldots a^m_{\left|C_m\right|}}_{C_m}\left| \underbrace{x^1_{1}  \ldots  x^1_{\left|C_1\right|}}_{C_1}
\underbrace{x^2_{1}  \ldots  x^2_{\left|C_2\right|}}_{C_2}
\ldots \underbrace{x^m_{1}  \ldots  x^m_{\left|C_m\right|}}_{C_m}\right.\right),\ee
\end{widetext}
where $m = \left|\mathcal{C}\right|$, that gives the joint  probability of 
obtaining outcomes $a^i_{1}, \ldots ,a^i_{\left|C_i\right|}$ for each 
context $ C_i=\left\{x^i_{1},  \ldots ,  x^i_{\left|C_i\right|}\right\}.$
Notice that this extended global probability distribution is, in general, 
not equal to the probability distribution defined in Eq.
\eqref{eqglobal}, since the same measurement may appear in more than 
one 
context, and hence, in the list 
\be \underbrace{x^1_{1}  \ldots  x^1_{\left|C_1\right|}}_{C_1}
\underbrace{x^2_{1}  \ldots  x^2_{\left|C_2\right|}}_{C_2}
\ldots \underbrace{x^m_{1}  \ldots  x^m_{\left|C_m\right|}}_{C_m} \ee 
the same measurement may be repeated several times.

To make  definitions  in Eqs.\eqref{eqglobal} and 
\eqref{eqextendedglobal} equivalent in the case of nondisturbing behaviors, we demand that, if in  different 
contexts $C_{i_1}, C_{i_2}, \ldots , C_{i_l}$ there exist coincident measurements $x^{i_1}_{k_1}, x^{i_2}_{k_2}, \ldots , x^{i_l}_{k_l}$ ,
then the marginal probability distributions  for  $x^{i_1}_{k_1},x^{i_2}_{k_2}, \ldots , x^{i_l}_{k_l}$  are perfectly correlated.
It is then equivalent to  say that  $\mathrm{B}$ is a \textit{ noncontextual behavior}
 if there is a 
extended global probability distribution  satisfying this condition such that   the marginals for each context coincide with the probability distributions 
of $\mathrm{B}$.

Consider the example framed in Fig.\ref{fig:c3}. Traditionally, see 
Eq.\eqref{eq:ExampleNonDisturb}, one says that a nondisturbing 
behavior for 
 this scenario is noncontextual if there is a global probability 
distribution $p(abc\vert xyz)$ such that $p(ab|xy)= \sum_c p(abc\vert xyz)$
 and $p(bc|yz)=\sum_a p(abc\vert xyz)$. An extended global probability distribution is a probability distribution
 $p(ab^1b^2c|xy^1y^2z)$ such that \begin{align} p(b^1b^2|y^1y^2) &=& \sum_{a,c} 
p(ab^1b^2c|xy^1y^2z) \nonumber \\
&=&\begin{cases}
                        1& \text{if} \ b^1=b^2\\
                        0&\text{otherwise}
                       \end{cases},\label{eq:ext_ex}\end{align}
 where $y^1$ and $y^2$ represent the two copies of measurement $y$, one for each context. Then 
we say that a behavior  is noncontextual if
 there is an extended global probability distribution satisfying 
condition \eqref{eq:ext_ex} such that 
\begin{align}
 p(ab^1|xy^1)= &\sum_{b^2,c} p(ab^1b^2c|xy^1y^2z)\\
p(b^2c|y^2z)=&\sum_{a,b^1} p(ab^1b^2c|xy^1y^2z).
\end{align}
For nondisturbing 
behaviors, these two notions
 of noncontextualtiy are equivalent~\cite{ADO18}.

To define noncontextuality in a scenario where the nondisturbance property does not hold, we adopt the strategy of Refs. \cite{KDL15,DK16,DKC16,DK17,DCK17}. That is to say,
we relax 
 the requirement that marginals for variables $x^{i_1}_{k_1},x^{i_2}_{k_2},\ldots ,x^{i_l}_{k_l}$ must be
perfectly correlated when they represent the same measurement. Instead, we require that the probability of 
$x^{i_1}_{k_1},x^{i_2}_{k_2},\ldots , x^{i_l}_{k_l}$
being equal is the maximum allowed by the individual probability distributions of each $x^{i_l}_{k_l}$.
 
 We say that a behavior has a \emph{maximally noncontextual description} if there is an extended global distribution \eqref{eqextendedglobal} such that 
the distribution of each context is obtained as a marginal and such that if  
$x^{i_1}_{k_1},x^{i_2}_{k_2},\ldots , x^{i_l}_{k_l}$
represent the same measurement, the joint marginal distribution for $x^{i_1}_{k_1},x^{i_2}_{k_2}, \ldots , x^{i_l}_{k_l}$  is such that 
\be p\left(x^{i_1}_{k_1}=\ldots = x^{i_l}_{k_l}\right)=\sum_{a} p\left(a\ldots  a\left|x^{i_1}_{k_1}\ldots  x^{i_l}_{k_l}\right.\right) \ee
is the maximum consistent with the  marginal distributions $p\left(a^{i_j}_{k_j}\left|x^{i_j}_{k_j}\right.\right)$.
In plain English, a behavior is noncontextual in the extended sense if there is an extended global distribution that gives the correct 
marginal in each context and that maximizes the probability of $x^{i_1}_{k_1},x^{i_2}_{k_2},\ldots , x^{i_l}_{k_l}$ being equal if they 
represent the same measurement in different contexts.

Given 
$\{ x_{i_1}^{k_1},x_{i_2}^{k_2},\ldots, 
x_{i_l}^{k_l}\}$  representing the same 
measurement, we call a distribution 
\be p\left(a_{i_1}^{k_1}a_{i_2}^{k_2}\ldots  
a_{i_l}^{k_l}\left|x_{i_1}^{k_1}x_{i_2}^{k_2}\ldots  
x_{i_l}^{k_l}\right.\right) \ee
that gives the correct marginals 
$p\left(a_{i_j}^{k_j}\left|x_{i_j}^{k_j}\right.\right)$ a \emph{coupling} 
for 
$x_{i_1}^{k_1},x_{i_2}^{k_2},\ldots , x_{i_l}^{k_l}$. We say that such a 
coupling is \emph{maximal}
if $p\left(x_{i_1}^{k_1}=x_{i_2}^{k_2}=\ldots = x_{i_l}^{k_l}\right)$ 
achieves the maximum value consistent with the marginals
$p\left(a_{i_j}^{k_j}\left|x_{i_j}^{k_j}\right.\right)$.

Although maximal couplings always exist, as shown in Ref. \cite{ADO18}, there is 
no guarantee that they are 
unique. Nonetheless, there are specific 
scenarios -- such as scenarios with two variables with any
number of outcomes and three variables each of which with two 
outcomes-- in which one can guarantee that this is indeed the case~\cite{ADO18}.  

Coming back once again to the example presented in Fig. \ref{fig:c3}, we say that a 
behavior is noncontextual in the extended sense if there is an extended global 
distribution 
\be p\left(ab^1b^2c \vert xy^1y^2z\right) \ee
such that 
\begin{align}
  p\left(ab^1\vert xy^1\right) =&\sum_{b^2,c} p\left(ab^1b^2c \vert xy^1y^2z\right)\\
 p\left(b^2c \vert y^2z\right)=&\sum_{a,b^1} p\left(ab^1b^2c \vert xy^1y^2z\right)
\end{align}
and 
\begin{align}
 p\left(y^1=y^2\right) & \coloneqq  \sum_{b}p\left(bb\vert y^1y^2\right) \nonumber \\
  & = \sum_{a,b,c} 
p\left(abbc\vert xy^1y^2z\right)
\end{align}
is maximal with respect to the marginals 
\begin{align}
 p\left(b^1\vert y^1\right)= &\sum_{a} p\left(ab^1 \vert xy^1\right)\\
p\left(b^2 \vert y^2\right)=&\sum_{c} p\left(b^2c\vert y^2z\right).
\end{align}


\subsection{Extended compatibility scenario}
\label{subsec:Extended_Comp_Hyper}

To build a toolbox for extended contextuality, we associate to any scenario $\Gamma =\left(X, 
\mathcal{C}, O \right)$ an extended scenario \be \Upsilon \coloneqq\left(\mathscr{X}, 
\mathscr{C}, O \right), \label{eq:DefExtendedScenario} \ee constructed in the 
following way: to each vertex $x \in X$, let $C_{i_1}, \ldots , C_{i_l}$ be all 
contexts containing it. The set  $\mathscr{X}$  consists of measurements denoted by
$x^{i_1}, \ldots , x^{i_l}$, which represent different copies of the measurement $x$, 
one for each context containing it. 
For each $x \in X$ the set $\left\{x^{i_1}, \ldots , x^{i_l}\right\}$ belongs to $\mathscr{C}$.
The other contexts in $\mathscr{C}$ are in one-to-one correspondence
with the contexts in  $\mathcal{C}$:  each context $ C_i=\left\{x_{1}, 
x_{2}, \ldots ,  x_{\left|C_i\right|}\right\}$ in $\mathcal{C}$ corresponds
the context $\left\{x_{1}^i, x_{2}^i, \ldots ,  
x_{\left|C_i\right|}^i\right\}$ in $\mathscr{C}$.
The 
\emph{extended compatibility hypergraph}
$\mathscr{H}$ of the scenario is the compatibility hypergraph of the extended scenario $\Upsilon$.

\begin{figure}[h!]
 \begin{subfigure}[b]{0.4\textwidth}
\includegraphics[scale=1]{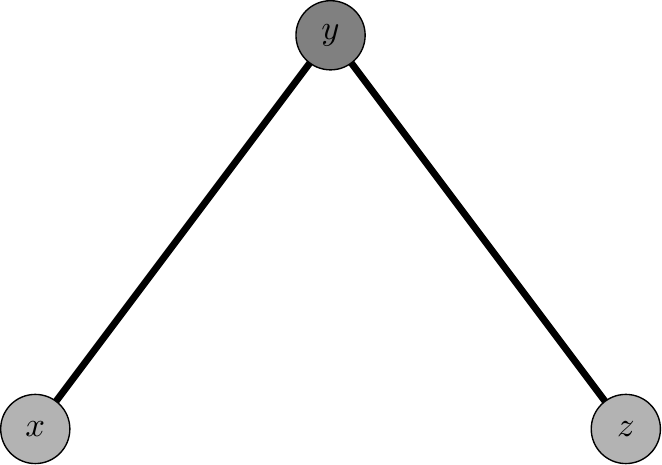}
\caption{ }
\end{subfigure}
\quad \qquad \quad
 \begin{subfigure}[b]{0.4\textwidth}
\includegraphics[scale=1]{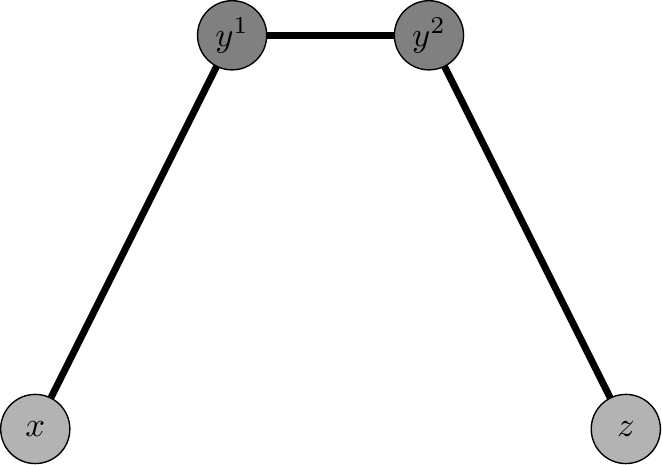}
\caption{  }
 
 \end{subfigure}
 \caption{\footnotesize{ (a)  Compatibility hypergraph $\mathrm{H}$ of  
Fig. \ref{fig:c3}.
 (b) The extended compatibility hypergraph $\mathscr{H}$ of $\mathrm{H}$.}}
  \label{fig:path_ext}
\end{figure}

 Fig. \ref{fig:path_ext} illustrates the extended compatibility hypergraph of the scenario defined in Fig. \ref{fig:c3}.
 In this scenario we have three measurements $x,y,z$ and two contexts, 
$C_1=\left\{x,y\right\}$ and $C_2=\left\{y,z\right\}$.
 Measurement $x$ belongs only to context $C_1$, measurement $z$ belongs only
 to context  $C_2$, while measurement $y$ belongs to both contexts. Hence,   $\mathscr{X}=\left\{x^1, y^1, y^2, z^2\right\}$.
 The contexts in $\mathscr{C}$ are $\left\{y^1, y^2\right\}$,  $\left\{x^1, y^1\right\}$
 and $ \left\{y^2, z^2\right\}$, the last two being the ones  corresponding 
to those of $\mathcal{C}$. Since $x$ and $z$ have only one copy in $\mathscr{X}$ we 
continue denoting these measurements simply by $x$ and $z$.

Given a behavior $\mathrm{B}$ for $\Gamma$,  we construct an \emph{extended behavior} $\mathscr{B}$
for  $\mathrm{B}$ in the following way:
for each context $\left\{x_{1}^i x_{2}^i \ldots  x_{\left|C_i\right|}^i\right\}$ of 
$\mathscr{C}$ corresponding to context $C_i=\left\{x_1 x_2 \ldots  
x_{\left|C_i\right|}\right\}$ of $\mathcal{C}$ 
the probability distribution  assigned by behavior $\mathscr{B}$ is equal to the probability distribution assigned to $C_i$ via the original behavior 
$\mathrm{B}$;
 for context  $x^{i_1}, \ldots , x^{i_l}$ of $\mathscr{C}$ corresponding to the different copies of a measurement $x \in X$, the probability distribution 
 assigned by behavior $\mathscr{B}$ is any maximal coupling for the variables $x^{i_1}, \ldots , x^{i_l}$.
  Since, in general, maximal couplings are not unique, $\mathscr{B}$ will 
also not be unique.

In the example of Fig.~\ref{fig:c3}, a behavior $B$ corresponds to two probability distributions $p\left(ab \vert xy\right)$ 
 and $p\left(bc \vert yz\right)$. An extended behavior $\mathscr{B}$ corresponds to three probability distributions $p\left(ab^1 \vert xy^1\right)$,
 $p\left(b^2c \vert y^2z\right)$ and $p\left(b^1b^2 \vert y^1y^2\right)$, such that the distribution for $xy^1$ is the same
 as the distribution for $xy$, 
 the distribution for $y^2z$ is the same as the distribution for $yz$ and $p\left(b^1b^2 \vert y^1y^2\right)$ maximizes the probability of
 $y^1$ and $y^2$ being equal given the  marginals $p\left(b^1\vert y^1\right)$ and $p\left(b^2 \vert y^2\right)$.

In Ref. \cite{DK17}, the authors define the notion of multimaximal coupling and use this notion to give a different  definition of extended contextuality. 
In this work we prefer to 
adopt maximal couplings as our starting point and by doing so we point out and discuss in details the differences between both approaches in Sec. \ref{sec:couplings}. We notice, however, that the tools used here apply to both cases, and, more generally,
to any kind of coupling one imposes on the sets $x^{i_1}, \ldots , x^{i_l}$ of $\mathscr{C}$. Physical constraints  shall ultimately decide which couplings are more meaningful.
Once the relevant couplings are defined, the notion of extended behavior will follow 
analogously and 
the mathematics from this step forward is exactly the same in all situations.

 With these concepts in hands, we can rewrite the definition of extended contextuality 
as the following theorem:

\begin{thm}
A behavior $\mathrm{B}$ for $\Gamma$ has a \emph{maximally noncontextual description} 
 if, and only if, there is an extended behavior  $\mathscr{B}$
for  $\mathrm{B}$ which is noncontextual in the traditional sense with respect to the extended  scenario $\Upsilon$.
\label{thm:extended}
\end{thm}

Thus, the problem of deciding whether a behavior $\mathrm{B}$ is noncontextual in the extended sense is equivalent to the problem of 
finding an extended behavior $\mathscr{B}$ which is noncontextual in the extended scenario $\Upsilon$. 
Hence, the tools needed for  the characterization of extended contextuality in the contextuality-by-default framework are exactly the same tools used in the characterization of 
contextuality in the traditional definition, with the complication that the scenario under study is more complex and that one may have to look to 
several different extended behaviors.

\begin{figure}[h!]
\begin{subfigure}[b]{0.4\textwidth}
\includegraphics[scale=1]{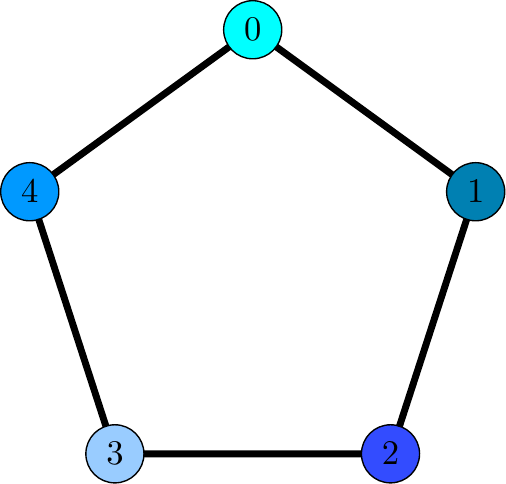}
\caption{}
\end{subfigure}
\begin{subfigure}[b]{0.4\textwidth}
\includegraphics[scale=1]{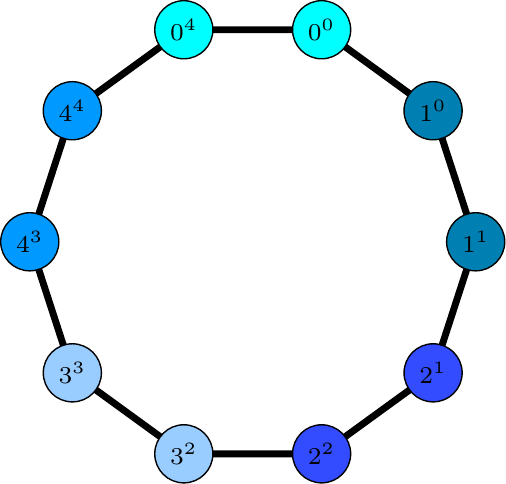}
\caption{}
\end{subfigure}
\caption{\footnotesize{(a) The compatibility hypergraph $\mathrm{H}$ of 
the $5$-cycle 
scenario, which consists of five measurements $0, \ldots , 4$ and five 
contexts $\left\{i,i+1\right\}$, $i=0, \ldots, 4$, the sum being  taken 
$\mod 5$. (b)  The extended compatibility hypergraph $\mathscr{H}$ 
of the $5$-cycle scenario, which is a $10$-cycle with vertices $i^{i-1}, 
i^{i}$ and egdes $\left\{i^i, (i+1)^i\right\}, \left\{i^i, i^{i-1} 
\right\}, \ i=0, \ldots , 4$. }}
\label{fig:ncycle}
\end{figure}

This gives, as corollary,  a complete characterization of extended contextuality for 
the $n$-cycle scenario \cite{KDL15,KD16,ADO18}.
In the $n$-cycle scenario, $X=\left\{0,\ldots , n-1\right\}$ and two 
measurements $i$ and $j$ are compatible iff $j=i+1 \mod n$. The 
corresponding hypergraph $\mathrm{H}$ is the cycle $\mathrm{C}_n$ with $n$ vertices. The 
extended hypergraph $\mathscr{H}$ is a  $2n$-cycle, with vertices 
$i^i, i^{i+1}$ and egdes $\left\{i^i, 
(i+1)^i\right\}, \left\{i^i, i^{i-1} \right\}, \ i=0, \ldots , n-1$ (see 
Fig. \ref{fig:ncycle}).

\begin{cor}
A behavior $B$ for the $n$-cycle scenario is noncontextual in the extended sense iff
\begin{align}
s\left(\left\langle {i}^i(i+1)^i\right\rangle, 1-\left|\left\langle 
i^{i}\right\rangle - \left\langle i^{i-1}\right|\right\rangle\right)_{i=0, \ldots 
, n-1} \nonumber\\ \hspace{-3em} \leq 2n-2,\label{eq:ncycle_ineq}
\end{align}
where
\be s\left(z_1, \ldots, z_k\right) = \max_{\gamma_i = \pm 1, \prod_i 
\gamma_i=-1 } \sum_{i=1}^k \gamma_i z_i. \label{eq:ncycle}\ee
\end{cor}

 In fact,  the extended behavior $\mathscr{b}$ is unique and, as shown in 
Ref. \cite{KDL15}, for every context $\left\{i^{i-1}, i^{i}\right\}$ corresponding to $i \in X$  we have that the  maximal coupling satisfy:
\be  \left\langle i^{i-1}i^{i} \right\rangle=1-\left|\left\langle i^{i-1}\right\rangle - \left\langle i^i\right\rangle\right|.\ee
As shown in Ref. \cite{AQBTC13}, 
Eq.~\eqref{eq:ncycle_ineq} is a necessary and sufficient condition 
for membership in the noncontextual set of the scenario defined by $\mathrm{C}_{2n}$ and the result follows from
Thm. \ref{thm:extended}.

\section{Necessary and Sufficient conditions for extended contextuality}
\label{sec:tools}

In this section we list several tools developed for the characterization of traditional contextuality. As a consequence of Thm.
\ref{thm:extended}, these tools  can also be applied directly
to the extended scenario to characterize extended contextuality in the contextuality-by-default framework.

\subsection{Testing Noncontextuality with Linear Programming}

Noncontextuality of a behavior $\mathrm{B}$ is equivalent~\cite{ACTA18} to the 
existence of a set of variables $\Lambda$ and deterministic probability distributions
$p\left(a_i \vert x_i, \lambda\right)$ for each $x_i \in X$ and $\lambda \in \Lambda$ such that 
\begin{align}
p\left(a_1, \ldots,a_{\left|C\right|} \vert x_1, \ldots,x_{\left|C\right|}\right) & \nonumber\\ = &\sum_{\lambda} p(\lambda) \prod_{i=1}^{\left|C\right|} p\left(a_i \vert x_i,\lambda\right).\label{eq:dec}
\end{align}
It is possible to show that it suffices to consider a set $\Lambda$ with the same number of elements as the extremal points of the noncontextual set~\cite{DSW18}.

The most general way of deciding whether a behavior can be written in the form 
\eqref{eq:dec} is through a linear program (LP) formulation \cite{DK16}. 
Representing each probability distribution as a vector, Eq.\eqref{eq:dec} can be 
written succinctly as \be \mathrm{B}=A \cdot \bm{\lambda}, \ee with $\bm{\lambda}$ 
being a vector with components 
\be \lambda_i=p(\lambda=i)\ee and $A$ being a matrix  whose columns are the deterministic distributions 
\be \prod_{j=1}^{\left|C\right|} p\left(a_j \vert x_j,\lambda_i\right), \ee
that is, the columns of $A$ are the extremal points of the noncontextual set. Hence, checking whether $\mathrm{B}$ 
is noncontextual amounts to solve a simple feasibility problem written as 
the following LP :
\begin{eqnarray}
\min_{\lambda \in \mathds{R}^m} & & \quad \quad \mathbf{v} \cdot \bm{\lambda} 
\nonumber  \\ 
\textrm{subject to } & &  \quad \mathrm{B}=A \cdot \bm{\lambda}  \\ \nonumber
& & \quad \lambda_i \geq 0 \\ \nonumber 
& & \quad \sum_{i}\lambda_i=1,
\end{eqnarray}
where $\mathbf{v}$ represents an arbitrary vector with the same dimension $m=\vert 
O\vert^{\vert X\vert } $ as the vector representing the  variable $\bm{\lambda}$.

As a consequence of Thm. \ref{thm:extended}, we have:

\begin{thm}
A behavior $\mathrm{B}$ is noncontextual in the extended sense if there is an extended behavior $\mathscr{B}$ in the extended scenario such that the following linear program 
\begin{eqnarray}
\min_{\lambda \in \mathds{R}^m} & & \quad \quad \mathbf{v} \cdot \bm{\lambda}  \\ \nonumber
\textrm{subject \  to } & &  \quad \mathscr{B}=\mathscr{A} \cdot \bm{\lambda}  \\ \nonumber
& & \quad \lambda_i \geq 0 \\ \nonumber 
& & \quad \sum_{i}\lambda_i=1,
\end{eqnarray}
is feasible, where $\mathscr{A}$ is a matrix whose columns are products of extremal 
points of the noncontextual polytope of the extended scenario.
\end{thm}

\subsection{Noncontextuality Inequalities}

In a scenario $\Gamma=(X,\mathcal{C},O)$, given  $\gamma_{s\vert C}$ and
 $ \mathrm{b} \in \mathds{R}$ we say that the linear inequality 
 \be \sum_{s \in O^{\left|C\right|}, C \in \mathcal{C} } \gamma_{s\vert C} p\left(s\vert C\right) \leq \mathrm{b} , \ee
  is a \emph{noncontextuality inequality}  if it
 is satisfied for all $\mathrm{B} \in  \mathcal{NC}\left(\Gamma\right)$.

 Every noncontextuality inequality gives rise to a necessary condition for noncontextuality in 
the corresponding scenario. The noncontextual set is a polytope and
 hence it is characterized by a finite number of noncontextuality 
inequalities~\cite{Amaral14}. The complete characterization of the noncontextual set
 requires one to find all facet defining inequalities and this is in general a hard problem, specially in the extended framework where 
 the number of measurements is large. Nevertheless, there are algorithms that can list all the facet-defining inequalities for small graphs.

 For scenarios where the contexts have at most two measurements and each measurement 
has two outcomes, we can explore the connection between the 
noncontextual set and  the \emph{cut polytope} 
$\mathrm{CUT}\left(\mathrm{H}\right)$ \cite{AII06, AT17} of the 
corresponding compatibility hypergraph $\mathrm{H}$. In this case $\mathrm{H}$ is 
nothing but a simple graph and, if the nondisturbance condition is satisfied, deriving 
necessary 
conditions for extended contextuality  
reduces to the traditional necessary conditions for noncontextuality.

A detailed 
discussion of this connection is presented in Ref. \cite{ADO18}. There the authors 
have proven that
the extended compatibility hypergraph $\mathscr{H}$ can be 
obtained from the compatibility hypergraph $\mathrm{H}$ combining graph operations 
known as 
\emph{triangular elimination}, \emph{vertex splitting} and \emph{edge 
contraction} \cite{BM86,AII08,BJRR14}. From valid inequalities for 
$\mathrm{CUT}\left(\mathrm{H}\right)$ it is possible to derive valid 
inequalities for any graph obtained from $\mathrm{H}$ using a sequence of 
such operations. In particular, for any valid inequality 
for $\mathrm{CUT}\left(\mathrm{H}\right)$ it is possible to derive valid inequalities 
for $\mathrm{CUT}\left(\mathscr{H}\right)$, among which there is one that 
reduces to the original inequality if the nondisturbance condition is 
satisfied. Hence, for every noncontextuality inequality in the traditional sense, we can derive a family of inequalities in the extended sense. Any violation of the inequalities obtained with this method implies that the extended behavior is contextual 
in the extended sense. 




\subsection{Noncontextuality Quantifiers}

Following the same reasoning we used above, extended noncontextuality of a 
behavior  can also be witnessed with noncontextuality quantifiers, \emph{i.e.} in 
order to define whether a given behavior is contextual we can also explore those 
functions associating each behavior $\mathrm{B}$ with a positive number 
$\mathrm{Q}\left(\mathrm{B}\right)$
such that $\mathrm{Q}\left(\mathrm{B}\right)=0$ if and only if $\mathrm{B}$ is 
noncontextual. 

Once again, as a consequence of Thm. \ref{thm:extended}, we have:

\begin{thm}
 A behavior $\mathrm{B}$ is noncontextual in the extended sense if and only if there is an extended behavior $\mathscr{B}$ such that $Q\left(\mathscr{B}\right)=0$
 for some contextuality quantifier $Q$. 
 \label{Thm:QuantifierFromThm1}
\end{thm}

In what follows we exhibit a number of monotones  of contextuality developed recently that can be used to witness extended contextuality in any scenario.

\subsubsection{Relative Entropy of Contextuality}

The \emph{relative entropy of 
 contextuality} of a behavior $B$ \cite{GHHHHJKW14, HGJKL15} is defined as
 \begin{align}\label{HorDist2}
E_{max}\left(B \right) & :=  \nonumber\\  \min _{B^{NC} \in \mathcal{NC}\left(\Gamma\right) }&\ \ 
\max _{C}\ \ 
 D_{\mathrm{KL}}\left(p(\cdot\vert C)  \middle\| p^{NC}(\cdot\vert C) \right),
\end{align}
where $p(\cdot\vert C)$ is the probability distribution given by the behavior $\mathrm{B}$ for context $C$, the minimum is taken over all noncontextual behaviors $B^{NC}=\left\{ p^{NC}(\cdot\vert C)\right\}$ and the maximum 
is taken over all
 contexts  $C \in \mathcal{C}$.

On the other hand, the \emph{uniform relative entropy of contextuality} of $B$
is defined as
\begin{align}\label{HorDist3}
E_{u}\left(B \right) & :=  \nonumber\\ \frac{1}{| \mathcal{C} |}  \min _{B^{NC}\in 
\mathcal{NC}\left(\Gamma\right) }& \ \ \sum _{C\in \mathcal{C}}\ 
D_{\mathrm{KL}}\left(p(\cdot\vert C) \middle\| p^{NC}(\cdot\vert C) \right),
\end{align}
  where the minimum is taken over all noncontextual behaviors $B^{NC}=\left\{ p^{NC}(\cdot\vert C)
\right\}$.

 Both $E_{max}$ and $E_{u}$ vanish if and only if $\mathrm{B}$ is noncontextual. In 
addition, while $E_{max}$ is a proper monotone in a resource theory 
 of contextuality with \emph{noncontextual wirings} as free-operations, 
the uniform relative entropy of contextuality is not. For more detailes, see Refs. 
\cite{GA17,ACTA18}.

\subsubsection{Distances}

In Refs. \cite{AT17,BAC18}, the authors define contextuality monotones  based on geometric distances, in contrast 
with the previous defined quantifiers which are based on entropic distances.
Let $D$ be any distance defined in real vector space  $\mathds{R}^K$. 
The $D$-\emph{contextuality distance} of a behavior $B$ is defined as
 \be \mathcal{D}\left(B\right):=\min _{\mathrm{B}^{NC}\in 
\mathcal{NC}\left(\Gamma\right) }  D\left(\mathrm{B} , \mathrm{B}^{NC} \right).
 \label{eqdefdist}\ee

We can also calculate the distance between the behaviors $B$ and $B^{NC}$ for each context $C$ and then average over the contexts. When the choice of context is uniform, we have the 
$D$-\emph{uniform contextuality distance} of a behavior $B$, defined as
\begin{align}
\mathcal{D}_u\left(B\right)& := \nonumber\\
\frac{1}{N}\min_{B^{NC}\in \mathcal{NC}\left(\Gamma\right) }& \sum_{C\in \mathcal{C}} D\left(p(\cdot\vert C) ,p^{NC}(\cdot\vert C) \right), \label{eqdefdist2}\end{align}
where $N=\left|\mathcal{C}\right|$ is the number of contexts in $\mathcal{C}$, $p(\cdot\vert C)$ is the probability distribution given by the behavior $\mathrm{B}$ for context $C$, and the minimum is taken over all noncontextual behaviors $B^{NC}=\left\{ p^{NC}(\cdot\vert C) \right\}$.

If we allow for non-uniform choices of context, the natural way of quantifying 
contextuality will be the $D$-\emph{max contextuality distance} of a behavior $B$, 
defined as
\begin{align} \mathcal{D}_{max}\left(B\right) &:= \nonumber \\  \min _{B^{NC}\in \mathcal{NC}\left(\Gamma\right) 
} & \max_{C}  D\left(p(\cdot\vert C), p^{NC}(\cdot\vert C) \right),  \label{eqdefdist3}\end{align}
where the minimum is taken over all noncontextual behaviors $B^{NC}=\left\{ p^{NC}(\cdot\vert C) \right\}$ and the maximum is taken over all
contexts  $C \in \mathcal{C}$.

The quantifiers $\mathcal{D}$, $\mathcal{D}_{u}$  and $\mathcal{D}_{max}$ vanish if and only if $\mathrm{B}$ is noncontextual.
The quantifier $\mathcal{D}_{max}$ is a proper monotone in a resource theory 
 of contextuality based on  \emph{noncontextual wirings} if the distance $d$ comes from 
a $\ell_p$-norm in $\mathds{R}^K$.
 If we choose the $\ell_1$-norm, $\mathcal{D}_{u}$ can be efficiently computed using linear programming \cite{AT17}. A detailed discussion of this quantifier for the special class of
Bell scenarios can be found in Ref. \cite{BAC18}. 
 
For the special class of $n$-cycle scenarios, the distance 
$\mathcal{D}_{u}$ defined with the 
$\ell_1$-norm is given by
\begin{widetext}
\be \mathcal{D}_u\left(B\right):=\frac{1}{N}\min_{B^{NC}\in 
\mathcal{NC}\left(\Gamma\right) } \sum_{C\in \mathcal{C}, a \in O^C} \left|p(a\vert C) 
-p^{NC}(a\vert C) \right|_{\ell_1}=\frac{1}{2}\max_s\left[s\left(\left\langle 
{i}^i(i+1)^i\right\rangle\right) - (n-2)\right],\ee
\end{widetext}
where $s$ was defined in equation \eqref{eq:ncycle}. This shows that  for $n$-cycle scenarios, the distance
$\mathcal{D}_u(B)$ defined by the $\ell_1$-norm is equal to the violation of the only noncontextual
inequality \eqref{eq:ncycle_ineq} violated by $B$.

In the extended $n$-cycle scenario, there is only one extended behavior $\mathscr{B}$ for each behavior $B$.
Hence, the quantifier  in the extended sense will also be given by
\begin{widetext}
\be \mathcal{D}_u\left(\mathscr{B}\right)=\frac{1}{2}\max_s\left[s\left(\left\langle {i}^i(i+1)^i\right\rangle, 1-\left\langle 
i^{i}\right\rangle - \left\langle i^{i-1}\right\rangle\right) - (2n-2)\right] \ee
\end{widetext}
and it will vanish if and only if $B$ is noncontextual in the extended sense.

\subsubsection{Contextual Fraction}

Broadly speaking, we may interpret the contextual fraction of a behavior as a 
contextuality quantifier based on the intuitive notion of which \emph{fraction} of it admits a noncontextual description. Such measure 
was introduced in Refs. \cite{AB11, ADLPBC12}, and several properties of this 
quantifier were further discussed in Ref. \cite{ABM17}. 

More specifically,
the \emph{contextual fraction} of a behavior $B$ is defined as
\be
\label{eq:cont_frac}
\mathcal{CF}\left(B\right)= \min \left\{\lambda \left|B=  \lambda B' + \left(1-\lambda\right)B^{NC}\right.\right\},
\ee
where $B^{NC}$ is an arbitrary noncontextual behavior. The contextual fraction vanishes if and only if $\mathrm{B}$ is noncontextual and it can be efficiently computed
using linear programming.

In the $n$-cycle scenario, each behavior $B$ violates only one facet-defining inequality 
\eqref{eq:ncycle_ineq}. This implies that we can write
\be B=  \mathcal{CF}\left(B\right) B' + \left(1-\mathcal{CF}\left(B\right)\right)B^{NC} \ee
where $B^{NC}$ is a noncontextual behavior saturating the inequality and $B'$ is the only contextual behavior that maximally violates the inequality. The linearity of the noncontextuality inequalities in turn implies that $B$ will violated this inequality by $\mathcal{CF}\left(B\right)$ times the violation obtained with $B'$. Hence,
we also have that 
\begin{widetext}
\be \mathcal{CF}\left(B\right)=\frac{1}{n}\max_s\left[s\left(\left\langle {i}^i(i+1)^i\right\rangle, 1-\left\langle 
i^{i}\right\rangle - \left\langle i^{i-1}\right\rangle\right) - (n-2)\right]. 
\label{Eq:CFNCycleNormal} \ee
\end{widetext}

Clearly, and analogously to what we have done for the $D$-uniform contextuality
distance, Eq.~\eqref{Eq:CFNCycleNormal} will also hold true in the extended 
scenario. 


\subsubsection{Negativity of global quasidistributions}

 A quasiprobability distribution is a set of real numbers $p_i$ such  that $ \sum_{i} p_i=1.$ 
If we relax the restriction that $p\left(a_1a_2 \ldots a_{n}\vert x_1 x_2 ... x_{\vert X \vert}\right)$ be a probability distribution 
and require only that it is be a quasiprobability distribution, then every nondisturbing behavior has a global quasiprobability distribution consistent with it \cite{AB11}.
Noncontextuality can be characterized in terms of these quasiprobability distributions: a behavior in noncontextual if and only if there is such global quasiprobability distribution
with  all $p\left(a_1a_2 \ldots a_{n}\vert x_1 x_2 ... x_{\vert X \vert}\right)\geq 0.$

We can also use these distributions to derive a contextuality quantifier 
\be \mathrm{N}\left(\mathrm{B}\right)= \min \sum_ {a, \mathrm{C}} \left|p\left(a_1a_2 \ldots a_{\left|X\right|}\vert x_1 x_2 ... x_{\vert X \vert}\right)\right|- 1\ee
where the minimum is taken over all global quasiprobability distributions consistent with $\mathrm{B}$ and the sum is taken over all contexts $\mathrm{C}=\left\{x_1, x_2, \ldots,  x_{\vert X \vert}\right\}$ and all 
possible outcomes $a= \left(a_1a_2 \ldots a_{\left|X\right|}\right) \in O^{\mathrm{C}}$. We have that $\mathrm{N}\left(\mathrm{B}\right)=0$ if and only if there is a global probability
distribution consistent with $\mathrm{B}$, and hence $\mathrm{N}$ is indeed a proper contextuality quantifier. It can be calculated using linear programming, since the sum can be formulated as
\begin{eqnarray}
\label{NLLP1}
\min_{\mathbf{t}  \in \mathds{R}^n, \lambda \in \mathds{R}^m} & & \langle \mathbf{1}_n, \mathbf{t} \rangle   \\ \nonumber
\mathrm{subjected \ to} & & \quad -\mathbf{t}  \leq \mathrm{B} - A \cdot \lambda \leq \mathbf{t}  \\ \nonumber
 & & \quad \sum_{i}\lambda_i=1, \\ \nonumber
\end{eqnarray}
where $A$ is the matrix whose columns are the extremal points of the noncontextual polytope.

As a consequence of Thm.~ \ref{thm:extended}, $B$ is noncontextual in the extended sense if and only if there is an extended behavior $\mathscr{B}$ for $B$ such that $\mathrm{N}\left(\mathscr{B}\right)=0$. For  $n$-cycle scenarios $\mathrm{N}\left(\mathscr{B}\right)$
is given by \cite{DK16}
\begin{widetext}
\be \mathcal{N}\left(\mathscr{B}\right)=\frac{1}{2}\max_s\left[s\left(\left\langle {i}^i(i+1)^i\right\rangle, 1-\left\langle 
i^{i}\right\rangle - \left\langle i^{i-1}\right\rangle\right) - (2n-2)\right].\ee
\end{widetext}

\subsection{Difference between maximal couplings and extended global distributions}

In Refs. \cite{DK17, KD16} the authors define a specific  contextuality quantifier for the contextuality-by-default approach based on the fact that no extended  global probability distribution
consistent with   the original behavior $\mathrm{B}$ can give a maximal coupling for  the copies  of all  measurements $x \in X$ if $\mathrm{B}$ is contextual.
Let $x^1, \ldots , x^n$ be the different copies of a measurement $x \in X$. We define
\be \mathrm{\mu}\left(\mathrm{x}\right)= \max p\left(x^1=x^2= \ldots = x^n\right) \ee
where the maximum is taken over all couplings of $x^1, \ldots , x^n$, that is, $\mu(x)$ gives the probability  of $x^1, \ldots, x^n$ being equal according to any maximal coupling for $x^1, \ldots, x^n$.

Given an extended global distribution $q$ we compute the probability of $x^1, \ldots, x^n$ being equal according to $q$:
\be \mathrm{m}^q\left(\mathrm{x}\right)=  q\left(x^1=x^2= \ldots = x^n\right). \ee

Combining these two quantities, we define 
\be \mathrm{M}_u\left(\mathrm{B}\right):= \sum_{x \in X} \mathrm{\mu}\left(\mathrm{x}\right) -  \max_q \sum_{x \in X} \mathrm{m}^q\left(\mathrm{x}\right),\ee
where the maximum is taken over all extended global distributions $q$ consistent with $\mathrm{B}$. Notice that $\mathrm{M}_u\left(\mathrm{B}\right)=0$ if and only if there is an extended global distribution consistent with some extended behavior $\mathscr{B}$ of $\mathrm{B}$, that is, if and only if $\mathrm{B}$ is noncontextual in the extended sense. This quantity can also be calculated using linear programming since it involves the maximization of the linear function 
\be \sum_x m^q(x) \ee
over the set of extended global distributions $q$ consistent with $B$ in each context $C$. 

It turns out that this quantifier is related to the the violation of the noncontextuality inequalities \eqref{eq:ncycle_ineq} for the $n$-cycle scenarios. In fact, it was shown in Ref. \cite{KD16} that 
\begin{widetext}
\be \mathrm{M}_u\left(\mathrm{B}\right) =\frac{1}{2} \max \left\{  s\left(\left\langle {i}^i(i+1)^i\right\rangle, 1-\left|\left\langle 
i^{i}\right\rangle - \left\langle i^{i-1}\right|\right\rangle\right)_{i=0, \ldots
, n-1} - (2n-2), 0\right\}.\ee
\end{widetext}
This proves that, for $n$-cycle scenarios,  the quantifiers $\mathrm{M}_u$, $N$, $\mathcal{CF}$ and $\mathcal{D}_u$ defined with the $\ell_1$-norm are all equivalent. This is also true for $\mathcal{D}_u$ defined with any $\ell_p$-norm,
as shown in Ref.~\cite{AT17}.

In references \cite{AG16,ACTA18} the authors show that the quantifiers $E_u$ and $D_u$ are not  monotones under the entire set of noncontextual wirings, as some preprocessings of the measurement labels $x_i$ may increase these quantifiers in an artificial way using only noncontextual resources. This shows that taking sums or means over the contexts generally do not lead to proper contextuality quantifiers when the entire class of noncontextual wirings is to be considered. Although we still lack a resource theory for contextuality that can be applied to the contextuality-by-default framework, we expect  the same problem to happen with the quantifier $\mathrm{M}_u$. It might be the useful then to take the maximum over measurements $x$ instead of the sum: 
\be \mathrm{M}\left(\mathrm{B}\right)= \min_q \max_{x \in X} \left[\mathrm{\mu}\left(\mathrm{x}\right) -  \mathrm{m}^q\left(\mathrm{x}\right)\right],\ee
where the minimum is taken over all extended probability distributions $q$ consistent with $\mathrm{B}$. Notice that we also have that  $\mathrm{M}\left(\mathrm{B}\right)=0$ if and only if $\mathrm{B}$ is noncontextual in the extended sense. Although possibly more suitable from the point of view of resource theories, this quantifier has the disadvantage of not being computed with linear programming.
Nevertheless, for the special case of  $n$-cycle scenario,  $\mathrm{M}_u$ and $\mathrm{M}$ coincide. In fact, it is always possible to find an extended global distribution that agrees with $\mathscr{B}$ 
for all contexts except one.

\section{Different Couplings}
\label{sec:couplings}

Following Refs. \cite{KDL15, DKC16}, we defined extended contextuality using the notion of maximal coupling. This kind of coupling has the property that 
a maximal coupling for a set of variables  is not necessarily a maximal coupling when we compute the marginal for a subset of these variables.
Hence, a noncontextual behavior $\mathrm{B}$ might become contextual in the extended sense when some measurements are ignored \cite{DK17}.

In Ref. \cite{DK17} the authors regard this property as a disadvantage of this kind of coupling and replace the constraint of maximal coupling by the constraint of 
\emph{multimaximal coupling}. A multimaximal coupling for a set of variables $x_1, \ldots, x_n$ is a maximal coupling for $x_1, \ldots, x_n$ such that if we compute the marginal distribution
for every subset of $x_1, \ldots, x_n$, the resulting distribution is also a maximal coupling for this subset.
The main disadvantage of this kind of constraint is that depending on the marginals for $x_1, \ldots, x_n$ this kind of coupling may not exist.

Our main motivation to study extended contextuality comes from the need of developing a formal structure for contextuality that can be applied to any experiment.
Hence, the multimaximal coupling constraint is not a good choice for this particular application since there is no guarantee that such a coupling exists for every experimental data. On the other hand, the fact that 
a maximal coupling for a set of variables  is not necessarily a maximal coupling when we compute the marginal for a subset of these variables can be intuitively explained in this situation.
Imagine that a measurement $x$ appears in four different contexts. In the extended scenario this measurement will correspond to four variables $x^1, x^2, x^3, x^4$. Suppose that the measurement of the 
first two contexts is perfect and the measurement of the last two contexts has a lot of errors.  It is possible that, even if you start with a setting that should in theory exhibit contextuality, the data will be 
noncontextual in the extended sense
if the measurements of the last two contexts are really bad. Hence, discarding these contexts will leave only the contexts where the measurement is perfect, and hence the data may become contextual.
Errors contribute to make the experimental behaviors noncontextual and it might be the case that discarding some measurements where the data is worse will make the behavior contextual in the smaller scenario.

We stress that the tools listed in Sec.~\ref{sec:tools} will work for any choice of coupling for the copies of the same measurement one  chooses. We believe that this choice may depend on the 
application and it is important to debate which choice is the most appropriate  in each situation. The generalization of contextuality provided by this framework 
can lead to a better comprehension of the data in contextuality experiments without unreal idealizations, but we believe that the kind of coupling one should impose needs to be further discussed.
Particularly, we should investigate if it is possible to justify the use of a specific coupling with experimental data. 
Nevertheless, once the choice is made the  notion of extended behavior will follow analogously and 
the mathematics from this step forward is exactly the same in all situations.

\section{Discussion}
\label{sec:discussion}
Apart from its primal importance in the foundations of quantum physics, 
contextuality has been discovered as a potential resource for quantum 
computing  \cite{Raussendorf13, HWVE14, DGBR14}, random number certification 
\cite{UZZWYDDK13}, and several other tasks in the particular case of Bell 
scenarios \cite{BCPSW13}.
 Within these both fundamental and applied perspectives, certifying 
contextuality  experimentally is undoubtedly an important primitive. It is 
then crucial  to develop a robust theoretical framework for contextuality 
that can be easily applied to real experiments. This should include the 
possibility of treating sets of measurements that do not satisfy the 
assumption of \emph{nondisturbance}, which will be hardly satisfied in  
experimental implementations~\cite{KDL15}.

It is in the pursuing of such an endeavor that the work we have presented 
here fits. On the one hand, inspired on the findings of the authors 
in Ref.~\cite{KDL15,DK16, DKC16, DK17,DCK17} and aware that a robust and easy-to-implement 
mathematical formalism should be 
established, we  further developed their extended definition of 
noncontextuality, rewriting it in graph-theoretical terms. It allowed us to explore geometrical aspects of the graph approach to 
contextuality to derive  conditions for extended contextuality 
that can be tested directly with  experimental data in any contextuality 
experiment and which reduce to traditional necessary conditions for 
noncontextuality if the nondisturbance condition is satisfied. In 
this sense, our proposal connects aspects of graph theory~\cite{DL97,BM86} 
with foundations of physics
~\cite{CSW10,CSW14,YO12,BBCP09,AFLS15} and experimental 
certifications~\cite{KDL15}. 

In addition, we also have centred our attention on how our formalism might be used in conjunction 
with known quantifiers in order to witness, in an alternative fashion, the notion of extended noncontextuality. In a nutshell, we have shown that the uniform relative entropy of contextuality, the uniform distance, as well as the contextual fraction and the negativity could all act as detectors, or witnesses for the extended notion of contextuality (see Thm.~\ref{Thm:QuantifierFromThm1}). Other common quantifiers as those based on robustness, say the robustness of contextuality \cite{HGJKL15,AT17}, could also have been approached and we believe similar results would also have been found. 

On another direction, it is known that the assumption of noncontextuality imposes non-trivial conditions on the Shannon entropies $\mathrm{H}\left(\mathrm{C}\right)$, and that these conditions can be written as linear inequalities \cite{Chaves13},  also known as entropic 
noncontextuality inequalities. Although in general they provide only necessary criteria for membership in the noncontextual set, the entropic framework reduces significantly the number of variables that have to be taken into account, an advantage that may be not only important but rather useful in the extended framework.  We have not explored this venue in here, but we would like to point out that this connection should be explored further though. 


 We believe that  the contextuality-by-default framework 
can lead to a better comprehension of the data in contextuality experiments, but  the restrictions of what kind of coupling one should impose in the different copies of the same measurement in 
different contexts needs to be further discussed.
It is imperative that future works  investigate whether it is possible to justify the use of a specific coupling given a certain set of experimental data. 

\begin{acknowledgments}
The authors thank  Jan-\AA{}ke Larsson, Ad\'an Cabello, Ehtibar N. Dzhafarov,  and Roberto Oliveira for 
valuable discussions.
Part of this work was done during the Post-doctoral Summer Program of Instituto de Matem\'atica Pura e Aplicada (IMPA) 2017. 
BA and CD thank IMPA for its support and 
hospitality.  Part of this work was done while BA was visiting Chapman University. BA thanks the university for its support and hospitality.
BA and CD acknowledges financial support from the Brazilian ministries and agencies MEC and 
MCTIC,  INCT-IQ, FAPEMIG, and CNPq. During the last stage of elaboration 
of this manuscript CD was also supported by a fellowship from the Grand Challenges 
Initiative at Chapman University. 
\end{acknowledgments}

\bibliography{biblio}

\end{document}